\documentclass[aps,pra,floatfix,showpacs,12pt,notitlepage]{revtex4-1}
\usepackage[english]{babel} 
\usepackage{amsmath} 
\usepackage[capitalise]{cleveref}
\usepackage{slashed} 
\usepackage{graphicx} 
\usepackage{color}
\usepackage{hepunits} 
\usepackage{bm} 
\usepackage{nicefrac}

\newcommand{\bra}[1]{\left\langle #1 \right\vert}
\newcommand{\ket}[1]{\left\vert #1 \right\rangle}
\newcommand{\diff}{\mathrm{d}}


\begin{document}
\allowdisplaybreaks

\title{Nonlinear single Compton scattering of an electron wave-packet}

\author{A. Angioi}
\affiliation{Max-Planck-Institut f\"ur
  Kernphysik, Saupfercheckweg 1, 69117 Heidelberg, Germany}

\author{F. Mackenroth}
\email{Currently at: Department of Applied Physics, Chalmers University of Technology, Gothenburg, Sweden}
\author{A. Di Piazza} \email{dipiazza@mpi-hd.mpg.de}
\affiliation{Max-Planck-Institut f\"ur
  Kernphysik, Saupfercheckweg 1, 69117 Heidelberg, Germany}

\date{\today}
\begin{abstract}
Nonlinear single Compton scattering has been thoroughly
investigated in the literature under the assumption that initially
the electron has a definite momentum. Here, we study a more general
initial state, and consider the electron as a wave-packet. In particular, we
investigate the energy spectrum of the emitted radiation
and show that in typical experimental situations 
some features of the spectra shown in previous works are almost completely
washed out. Moreover, we show that at comparable relative uncertainties,
the one in the momentum of the incoming electron has a larger 
impact on the photon spectra at a fixed observation direction than
the one on the laser frequency.
\end{abstract}

\maketitle

\section{Introduction}

According to classical electrodynamics a charged particle 
(an electron, for definiteness) accelerated by a
background electromagnetic field emits radiation~\cite{jackson}. In
the underlying quantum theory, QED, the radiation process is rather
described as the emission of photons by the electron
\cite{schroeder2005introduction,berestetskii1982quantum}.  Due to
energy-momentum conservation a free electron is stable and cannot emit
photons. The scattering of an electron with a single photon is known as (linear) Compton scattering. In
general, the simultaneous interaction of an electron with many photons is
suppressed by the appearance in the interaction probabilities of a corresponding power of the
fine-structure constant $\alpha_{QED} \approx 1/137 \ll 1$. However, if the
electron interacts with a coherent collection of photons, like those
in a laser beam, the effective coupling strength appearing in perturbative
expansions is not just $\alpha_{QED}$, but it also depends on the typical
amplitude and angular frequency of the laser field
\cite{ritus}. Qualitatively it is clear that a laser field
characterized by an amplitude $\mathcal{E}$ and by an angular
frequency $\omega$ is able to transfer to the electron (charge $e < 0$
and mass $m$) a number of photons of the order of
\begin{equation}
  \xi = \frac{|e|\mathcal{E}}{\omega m c},
\end{equation}
in the typical QED length $\lambda_C=\hbar/mc\approx 3.9\times
10^{-11}\;\text{cm}$ (Compton wavelength)
\cite{ritus,AntoninoRev,ehlotzky}. Thus, at $\xi\gtrsim 1$ the
probability for the electron of exchanging more than one photon with
the laser field is not suppressed and the laser-electron interaction
has to be taken into account exactly in the calculations. From a
classical point of view, the condition $\xi\gtrsim 1$ corresponds to
the onset of relativistic effects in the electron dynamics, which
render the latter nonlinear with respect to the laser amplitude. Now,
if the electron enters a plane-wave field with a four-momentum
$p^{\mu}=(\varepsilon/c,\bm{p})$, with
$\varepsilon=(m^2c^4+\lvert\bm{p}\rvert^2c^2)^{1/2}$, and in the process of photon
emission it absorbs $\xi$ laser photons, due to the Doppler effect,
the typical energy $\hbar\omega'$ of the emitted photon is of the
order of $\chi\varepsilon$, where
\begin{equation}
  \chi = \frac{(pk)}{m \omega}\, \frac{\mathcal{E}}{\mathcal{E}_{cr}},
  \label{eq:chi}
\end{equation}
with $k^\mu = (\omega/c, \bm{k})$ being the plane-wave's
four-wave-vector and $\mathcal{E}_{cr} = m^2c^3/\hbar |e|
\approx 1.3 \times 10^{16} \; \text{V}/\text{cm}$ being
the so-called ``critical field'' of QED~\cite{ritus,AntoninoRev}. The 
above estimate of the typical energy of the emitted photon
is valid for $\chi\lesssim 1$. A
constant and uniform electric field of the order of $\mathcal{E}_{cr}$
provides an $e^-$-$e^+$ pair with an energy comparable to its rest
energy $2 m c^2$ on a distance of the order of $\lambda_C$, such that
the QED vacuum becomes unstable in the presence of such a strong field
under $e^-$-$e^+$ pair creation~\cite{sauter, heiseul,
  gaugeinvvacumpol}. The parameter $\chi$ controls the importance of
photon recoil, which becomes essential at $\chi\gtrsim 1$. The emission 
of a single photon in the regime $\xi,\,\chi \gtrsim 1$ is known as 
nonlinear single Compton scattering and it
has been studied thoroughly in the literature
\cite{brown, goldman, nikiritus, ivanov, boca_nonlinear_2009, harveyheinzlilderton,
felix2010, boca_nonlinear_2011, felixnscs, seipt, krajkami, bocaal1, bocaal2,tobias, seipt2,
kraj2015}.

Although initial electron wave-packets have been considered in some studies about 
Thomson scattering~\cite{peatross}, and the general problem of the radiation
emitted by a classical distribution of charges is a well-known problem in the Free-Electron-Laser 
community~\cite{pellegrinireview}, in the study of nonlinear Compton scattering, 
to the knowledge of the authors, 
the initial state of the electron has been mostly taken
as having a definite momentum. In experiments, however, an
electron in a beam has some characteristic indeterminacy in the
momentum and is localized to some extent; motivated by this fact, we
will consider below that the electron is initially in a superposition
of states of different momenta, i.e., in a wave-packet, and, among
other aspects, we study whether it is possible to observe interference 
effects among different components of the wave-packet. We do this in the framework of 
strong-field QED within the Furry picture~\cite{furry,AntoninoRev, ritus, ehlotzky}.
In~\cite{corson1} a scalar QED calculation with an initial particle described by a wave-packet shows how
for nonlinear single Compton scattering
in a monochromatic plane-wave electromagnetic field the different components of the electron
wave-packet do not interfere. This result has been extended in~\cite{corson2}
to spinor QED in pulsed fields. We will show in the following a different derivation
of the same result, and in addition we will investigate in detail 
the effects of the initial electron's wave-packet on the emitted radiation.

Peak laser intensities have been recently increasing dramatically due to
the development of two techniques, Chirped Pulse
Amplification (CPA)~\cite{Strickland1985219} and Optical Parametric Chirped
Pulse Amplification (OPCPA)~\cite{opcpa}. All today's most intense lasers,
like Vulcan~\cite{vulcan}, Astra-Gemini~\cite{astra}, HERCULES~\cite{herc}, Berkeley Lab Laser Accelerator (BELLA)~\cite{BELLA}, and planned
ones, such as the Extreme Light Infrastructure~\cite{eli}, the High
Power Laser Energy Research facility~\cite{hiper}, APOLLON~\cite{apollon}, and the
Exawatt Center for Extreme Light Studies (XCELS)~\cite{xcels}, are
based on either one of these techniques.  Both CPA and OPCPA
generate, after the amplification of an initial pulse, an ultrashort
laser pulse; it is thus likely that this kind of pulses will be
adopted in experiments to probe the nonlinear QED regime. Thus, we will
consider the laser field in our calculations to be an ultrashort pulse.

To the current date, the record for the highest laser intensity ever
achieved is held by the HERCULES facility, that reached a peak intensity of $2 \times
10^{22}\; \text{W}/\text{cm}^2$ ($\xi \approx 70$ at $\hbar\omega=1.55\;\text{eV}$), 
and lasers with peak intensity $\xi \gtrsim 1$ are readily available in many
facilities. It is harder, however, to reach values of the 
parameter $\chi$ close to unity. Starting from \cref{eq:chi}, and
substituting the previously given definition of $\mathcal{E}_{cr}$,
one can write $\chi = \xi \; \hbar(p k)/m^2c^2$; the factor $\hbar(p k)/m^2c^2$ makes it
necessary, in order to have a $\chi$ close to unity at optical frequencies, to use
ultrarelativistic electrons (even for large values of $\xi\sim 100$).
Nowadays, ultrarelativistic electron beams can be conveniently produced also at laser
facilities via the wakefield acceleration technique~\cite{malka2008principles,leemansbeam}.

Although intense pulses are usually focused almost down to diffraction
limit, we will model them as plane waves. This approximation is
valid if the electron collides nearly head-on with the laser field and 
almost at the focus of the latter, provided that the transverse excursion of the electron is 
much smaller than the laser waist size, which occurs if $\xi mc^2 \ll
\varepsilon$~\cite{landau2013classical, berestetskii1982quantum}. Within the
plane-wave approximation, the approach based on the Furry picture
can be conveniently applied as the Dirac equation in a plane-wave field
can be solved exactly. Approximate solutions can be also found, however, for
a field of more complex structure like a Gaussian laser beam if the
conditions $\xi\gg 1$ and $\xi mc^2 \ll \varepsilon$ are fulfilled~\cite{ultra}.

In most of the numerical work performed to obtain the results in this paper, 
one of the main challenges is to perform integrals of highly oscillating functions;
typical quadrature schemes cannot be adopted, since they become more and more inaccurate
as the frequency of the oscillations of the integrand increases. 
Thus, we have used Filon's method~\cite{filon, iserles} to deal with this kind of integrals.
The basic idea behind it is to put an highly oscillating integral in the form $\int_\mathcal{I} \diff x\, f(x) e^{i a x}$,
where $f(x)$ is a smooth and sufficiently well behaved function, $a \gg 1$ is a constant, and
$\mathcal{I}$ is an interval in $\rm I\!R$; then divide $\mathcal{I}$ in some subintervals $\{\mathcal{I}_n, 
\, n \in \rm I\!N\}$, sufficiently small that in each of them the function $f(x)$ can be accurately approximated with a quadratic polynomial.
Then, in each subinterval the starting integrals are approximated
by a weighted sum of terms each having the form $\int_{\mathcal{I}_n} \diff x\, x^j e^{i a x}$, where
$j \in \{0,1,2 \}$, and each of these integrals can be evaluated analytically.
The advantage of this method is that the accuracy of the estimate increases
with increasing $a$.

This paper is organized as follows. In \cref{sec:theo}, we present the
general theory of the scattering of an electron in a superposition of
states with different momenta and a short intense laser pulse and we
show that interference effects among states with different momenta 
are not present. In \cref{sec:diffpz}, we study the
particular case of an electron wave-packet colliding head-on with a
laser pulse and of normally distributed longitudinal momentum, while
in \cref{sec:multi} we investigate the more general case where there
is also an indeterminacy on the transverse components of the
momentum. Through the rest of the article, natural units ($\hbar = c =
1$) are adopted, and the electromagnetic units used are such that the
QED coupling constant is $\alpha_{QED} = e^2 (\approx 1/137)$.

\section{Theory}\label{sec:theo}

In the computation of nonlinear single Compton scattering rates,
perturbative approaches with respect to the laser field can quickly
become impractical, when a sufficiently strong incoming
electromagnetic field is considered.  In fact, as we mentioned in the
introduction, for an incoming laser field such that $\xi
\gtrsim 1$ the exchange of many photons between the laser and the
electron becomes important and perturbative calculations up to
a very high order would be necessary. Typically, however, such intense
fields consist of an enormous number of coherent photons; this makes
it possible~\cite{berestetskii1982quantum} to neglect the quantum
nature of the background field and to treat it as a classical given
electromagnetic field.  By working within this approximation, one can split the
electromagnetic field four-vector potential into two parts: a
classical part, that accounts for the intense laser field, and a
quantized part, that accounts for all the other excitations of the
electromagnetic field, i.e., the radiation emitted by the
electron. After that, the electron-positron field is quantized by
taking into account exactly the background laser field. This is the
so-called Furry picture of QED~\cite{furry, berestetskii1982quantum},
which we mentioned in the introduction, and all the following calculations are performed within this
formalism.
\begin{figure}
  \centering
  \includegraphics[width= 0.4 \linewidth]{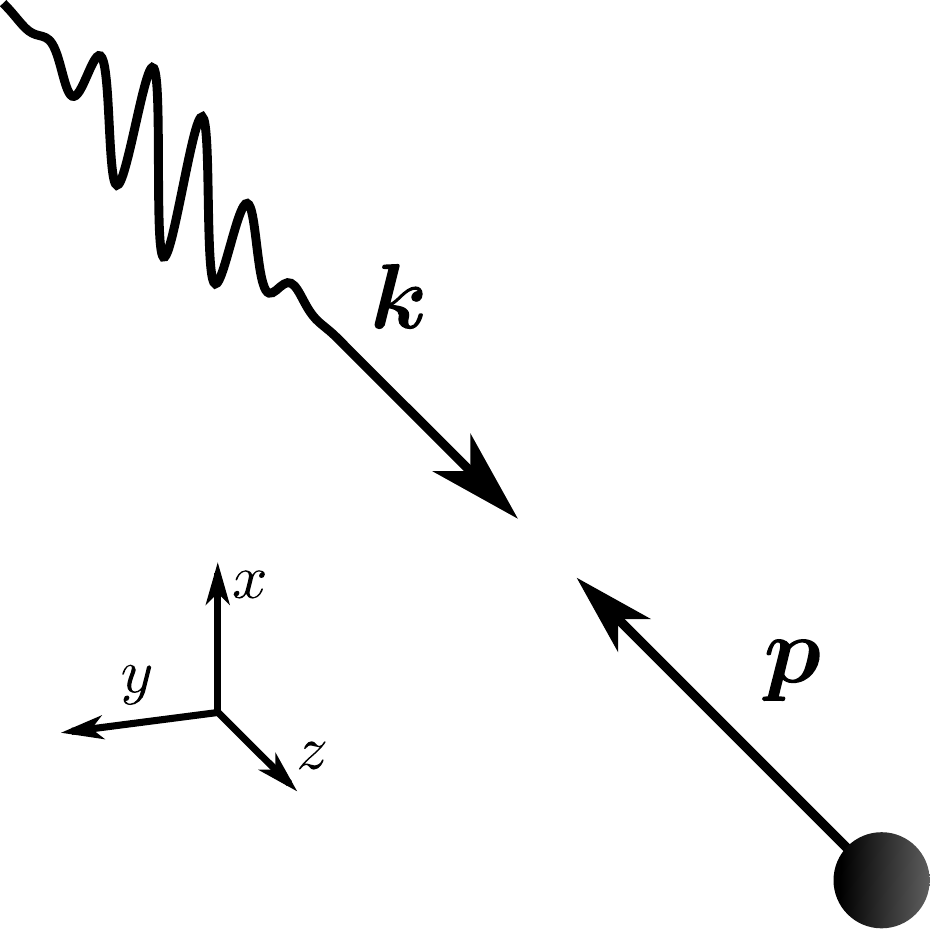}
  \caption{Representation of the choice of the employed frame of
    reference.}
  \label{fig:for}
\end{figure}

We assume that the incoming laser field is described by the
linearly-polarized plane-wave four-vector potential
\begin{equation}
  A^\mu(\eta) = \mathcal{A}^\mu \,\psi(\eta).
  \label{eq:vecpot}
\end{equation}
Here, $\mathcal{A}^\mu = (0,\bm{\mathcal{A}})$ is a constant four-vector, where $\bm{\mathcal{A}}$ defines the
laser polarization, with amplitude $\mathcal{A}=\mathcal{E}/\omega$ related to the peak laser's
intensity $I$ as $I=\omega^2\mathcal{A}^2/4\pi=\mathcal{E}^2/4\pi$, and $\psi(\eta)$ is a function of the laser phase $\eta=(kx)$ describing the shape of the plane wave and such that $\lvert \psi(\eta)\rvert \sim \lvert\diff\psi(\eta)/\diff\eta \rvert \le 1$. It is convenient to use a frame of reference in which one of the spatial axes (in our case the $z$ axis, for the sake of definiteness) is
directed along $\bm{k}$, and another one (without loss of generality,
we can choose $x$) is directed along the same direction as $\bm{\mathcal{A}}$ (see
\cref{fig:for}). Thereby, we have $\eta = \omega (t-z) = \omega
\varphi$, where $\varphi = t-z$.  It is also useful to introduce a
coordinate $T = (t+z)/2$, linearly independent of $\varphi$, $x$ and
$y$, and the quantities $\varphi$, $T$, $x$ and $y$ provide the
so-called light-cone coordinates of the space-time point $x^{\mu}$
(the factor of $1/2$ in the definition of $T$ is arbitrary and we have
chosen it in order for the Jacobian of the transformation from
Cartesian coordinates to light-cone ones to be unity). In the
following, we will define the $-$ (minus) contravariant component of
any four-vector $q^\mu$ to be $q^- = q^0 - q^3$.

\begin{figure}
  \centering
  \includegraphics{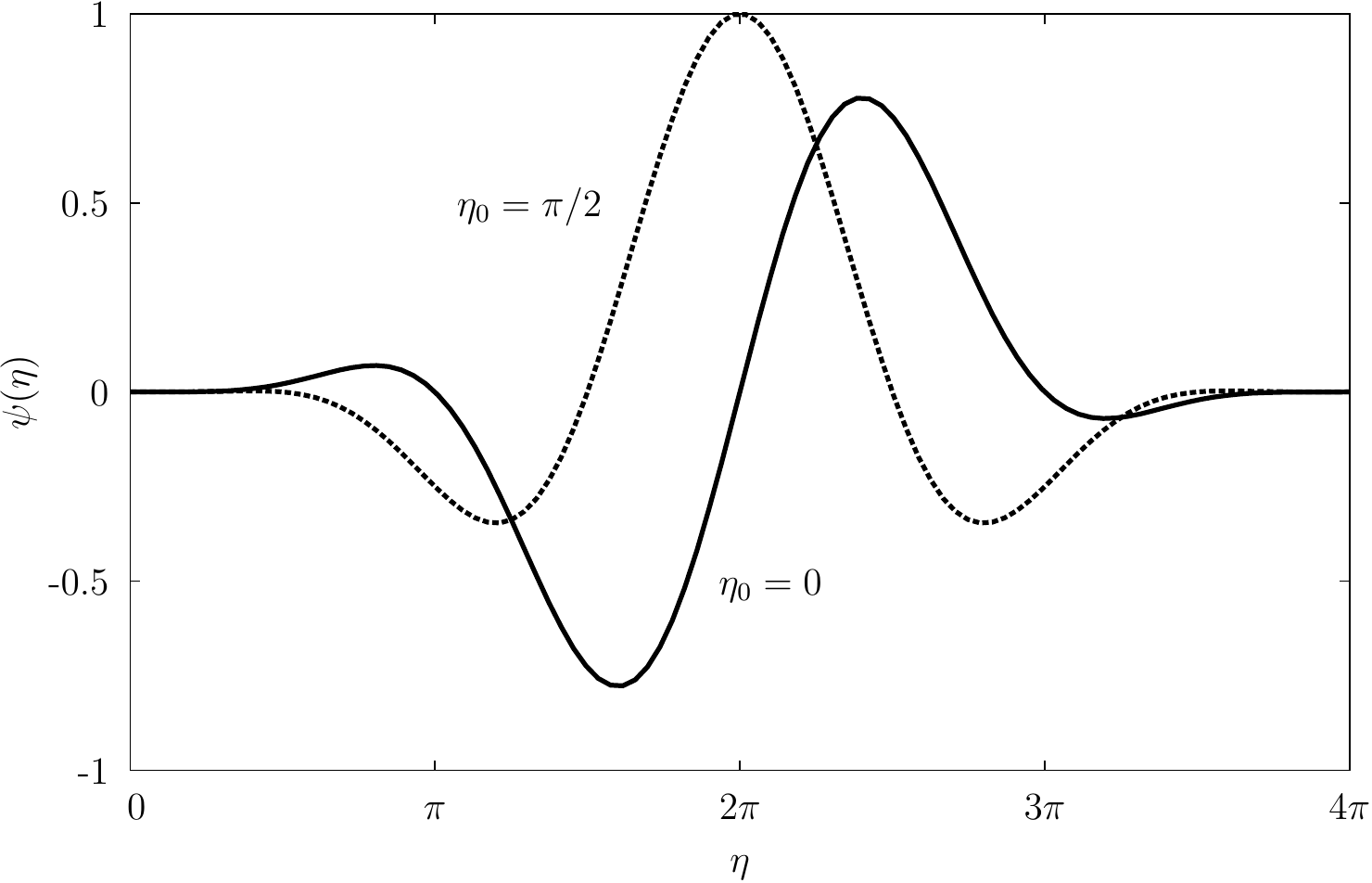}
  \caption{The function $\psi(\eta)$ for a two-cycle laser pulse ($n_C
    = 2$) and two choices of the carrier-envelope phase $\eta_0$. The
    solid curve corresponds to $\eta_0 = 0$, while the dotted one
    corresponds to $\eta_0 = \pi/2$.}
  \label{fig:shape}
\end{figure}

In the expression of $A^\mu(\eta)$ we have introduced the shape function
$\psi(\eta)$ in order to model short laser pulses; a typically
chosen~\cite{felixnscs} shape function $\psi(\eta)$ for this purpose
is (see \cref{fig:shape})
\begin{equation}
  \psi(\eta) = 
  \begin{cases}
    \sin^4\left( \frac{\eta}{2 n_C} \right) \sin(\eta + \eta_0)&
    \text{if } \eta \in
    [0,2\pi n_C], \\
    0 & \text{otherwise.}
  \end{cases}
\end{equation}
In this parametrization of the laser field we have introduced the
parameters $n_C$, the number of cycles contained in the laser pulse,
and $\eta_0$, the carrier-envelope phase (CEP) of the laser pulse.
In all the numerical calculations in the following, we will chose
$\eta_0 = 0$, $n_c = 2$, and $\omega = 1.55\; \eV$.

In the Furry picture, the states of the electron are described by the
solutions of the Dirac equation in the presence of the background
field; if this field is the plane wave $A^\mu(\eta)$, these solutions
(known as Volkov states) are given by~\cite{volkov,
  berestetskii1982quantum}
\begin{equation}
  \Psi_{p,\sigma} (x)  = \left[ 1 + \frac{e}{2(k p)} \slashed{k} \slashed{A}(\eta) \right] u_{p,\sigma} \,
  e^{-i p x - i \int_{-\infty}^{\eta} \left[ \frac{e}{(k p)} (pA(\eta')) -  \frac{e^2}{2(k p)} A^2(\eta')\right]\diff\eta'},
  \label{eq:volkovstates}
\end{equation}
where the slash on a four-vectorial quantity is a shorthand notation
for a contraction of that four-vector with the Dirac matrices
$\gamma^\mu$, that is, $\slashed{a} = \gamma^\mu a_\mu$ and
$u_{p,\sigma}$ is a positive-energy spinor solution of the free Dirac equation
$(\slashed{p} - m)\, u_{p,\sigma} = 0$, with $\overline{u}_{p,\sigma}
u_{p,\sigma} = 2m$
($\overline{u}_{p,\sigma}=u^{\dag}_{p,\sigma}\gamma^0$).  The Volkov
state $\Psi_{p,\sigma}(x)$ is characterized by the four-momentum
$p^\mu = (\varepsilon, \bm{p})$
($\varepsilon = \sqrt{m^2 + \lvert \bm{p} \rvert^2}$) and by the spin
quantum number $\sigma$ at $t\to-\infty$ (these are the so-called
Volkov in-states, although Volkov out-state only differ from the
in-ones by a phase independent of the coordinates). The Volkov states
in Eq. (\ref{eq:volkovstates}) are normalized as
\begin{equation}
  \int \diff^3 x \, \Psi^\dagger_{p',\sigma'}(x) \Psi_{p^,\sigma}(x) = (2\pi)^3\,(2 \varepsilon)\, \delta(\bm{p} - \bm{p}')\,\delta_{\sigma,\sigma'}.
\end{equation}

As we have mentioned in the introduction, we consider an electron
state which is a wave-packet made of a superposition of Volkov states
with different momenta and a given spin number $\sigma$:
\begin{equation}
  \Phi_\sigma(x) = \int \frac{\diff^3 p}{(2 \pi)^3 (2 \varepsilon)} \,\, \rho(\bm{p}) \,\Psi_{p,\sigma}(x).
  \label{eq:super}
\end{equation}
Here, $\rho(\bm{p})$ is a complex-valued, scalar weighting function;
in order for the state $\Phi_\sigma(x)$ to be normalized to unity,
$\rho(\bm{p})$ needs to be normalized in a covariant way as
\begin{equation}
  \int \frac{\diff^3 p}{(2 \pi)^3 (2\varepsilon)} \, \left\vert \rho(\bm{p}) 
  \right\vert^2 = 1.
\end{equation}

The leading order $S$-matrix element relative to the process of the
emission of a photon, with wave four-vector $k'^\mu = (\omega',
\bm{k}')$ and polarization four-vector $\epsilon'^\mu_l$, by an
electron in the initial state $\Phi_\sigma(x)$ is
\begin{equation}
  S_{fi} = -i e \sqrt{4 \pi} \int \diff^4 x \, \frac{\diff^3 p}{(2 \pi)^3 (2 \varepsilon)} \,\,
  \rho(\bm{p}) \, \overline{\Psi}_{p',\sigma'}(x)\, \slashed{\epsilon}'^*_l
  e^{i k' x} \, \Psi_{p,\sigma} (x).
  \label{eq:smatrix}
\end{equation}
We notice that among the space-time coordinates the integrand in \cref{eq:smatrix} depends
non-trivially only on $\varphi$, while on the other three space-time coordinates 
we have integrals that evaluate to three delta functions.  It is thus possible~\cite{felixnscs} to write
$S_{fi}$ in the form
\begin{equation}
  S_{fi} = -i e \sqrt{4 \pi}\, (2 \pi)^3 \int \frac{ \diff^3 p}{(2 \pi)^3  (2\varepsilon)} \, 
  \, \rho(\bm{p}) \,
  \left(\overline{u}_{p',\sigma'} M_{fi}u_{p,\sigma}\right) \,
  \delta^{(-,x,y)}\left( p - k' - p' \right);
\end{equation}
here, $\delta^{(-,x,y)}\left( p - k' - p' \right)$ is a three
dimensional Dirac delta that ensures the conservation of the three
contravariant components $-$, $x$ and $y$ of the total four-momentum
and
\begin{equation}
  M_{fi} = \slashed{\epsilon}'^*_l f_0 + e \left( \frac{\slashed{\mathcal{A}} \slashed{k} \slashed{\epsilon}'^*_l}{2 (k p')} + \frac{\slashed{\epsilon}'^*_l \slashed{k} \slashed{\mathcal{A}}}{2 (k p)} \right) f_1 
  - \frac{e^2 \mathcal{A}^2 (k \epsilon'^*_l) \slashed{k}}{2 (k p) (k p')} \, f_2, 
  \label{eq:emme}
\end{equation}
\vspace{-1.2em}
\begin{equation}
  f_j = \int_{-\infty}^{+ \infty}\diff\eta \, \psi^j(\eta) e^{i \int_{-\infty}^{\eta} \diff\eta'\, \left[ \alpha \psi(\eta')
      + \beta \psi^2(\eta') + \gamma \right]}.
  \label{eq:dinint}
\end{equation}
In \cref{eq:dinint} we have introduced the three
parameters~\cite{mackenroth2014quantum}
\begin{align}
  \alpha &= e \left[ \frac{(p' \mathcal{A})}{(k p')} - \frac{(p \mathcal{A})}{(k p)} \right], \\
  \beta &= -\frac{e^2 \mathcal{A}^2}{2} \, \frac{(k'k)}{(k p)(kp')}, \\
  \gamma &= \frac{(p k')}{(p' k)}.
  \label{eq:gamma}
\end{align}

In order to compute emission rates, it is necessary to calculate the
square modulus of $S_{fi}$:
\begin{multline}
  \lvert S_{fi}\rvert^2 = 4 \pi e^2 \int \frac{ \diff^3
    p}{(2\varepsilon)} \, \frac{ \diff^3
    \tilde{p}}{(2\tilde{\varepsilon})} \, \, \rho^*(\tilde{\bm{p}})
  \,\rho(\bm{p}) \left(\overline{u}_{p',\sigma'}
    M_{fi}u_{\tilde{p},\sigma}\right)^*
  \left(\overline{u}_{p',\sigma'} M_{fi}u_{p,\sigma}\right) \, \\
  \times\delta^{(-,x,y)}\left( p - k' - p' \right)\, \delta^{(-,x,y)}\left(
    \tilde{p} - k' - p' \right).
  \label{eq:squad}
\end{multline}
The integrations in \cref{eq:squad} are
along the components of $\bm{p}$ and $\tilde{\bm{p}}$ in Cartesian
coordinates, while one of the delta functions in \cref{eq:squad} is
expressed in terms of light-cone coordinates. An easy way to perform
these integrations is to change the measure for each momentum
integration from $\diff p_x \diff p_y \diff p_z = \diff^2 p_\perp
\diff p_z$ to $\diff^2 p_\perp \diff p^-$; the Jacobian one has to
insert for this transformation is $\varepsilon/p^-$.  Thus one can
start from \cref{eq:squad}, change the integration measure to $\diff^2
p_\perp \diff p^- \diff^2 \tilde{p}_\perp \diff \tilde{p}^-$, perform
the integrations in $\tilde{p}$ (that are just integrations of delta
functions), and change back the measure to $\diff^3 p$; this gives
\begin{equation}
  \lvert S_{fi}\rvert^2 = 4 \pi e^2\int \frac{ \diff^3 p}{(2 \varepsilon) (2 p^-)} \, 
  \lvert \rho(\bm{p}) \rvert^2 \,\,
  \lvert\overline{u}_{p',\sigma'} M_{fi}u_{p,\sigma}\rvert^2 \,\,
  \delta^{(-,x,y)}\left( p - k' - p' \right).
\end{equation}

The unpolarized emission rate is obtained by integrating $\lvert
S_{fi}\rvert^2$ over the electron's final momentum and on the
wave-vector of the emitted photon, and by summing over the final
electron spin and photon polarization, and averaging on the initial
electron spin
\cite{schroeder2005introduction,berestetskii1982quantum}:
\begin{multline}
  \diff W = \frac{\diff^3 k}{(2\pi)^3 (2\omega')} \,\int \frac{\diff^3
    p'}{(2\pi)^3 (2 \varepsilon')}\,
  \frac{ \diff^3 p}{(2 \varepsilon) (2 p^-)} \\
  4 \pi e^2\, \lvert \rho(\bm{p}) \rvert^2 \, \delta^{(-,x,y)}\left( p
    - k' - p' \right) \, \frac{1}{2} \sum_{\sigma, \,\sigma', \, l}
  \lvert\overline{u}_{p',\sigma'} M_{fi}u_{p,\sigma}\rvert^2.
\end{multline}
The integral on $\diff^3 p'$ can be readily evaluated with the same
change of integration measure previously mentioned. By writing
$\diff^3 k' = \omega'^2 \diff \omega' \diff \Omega'$ and remembering
that the emission rate and the energy emission rate are related by
$\diff E = \omega' \diff W$, it is possible to write the angular
differential emission rate as
\begin{equation}
  \frac{\diff E}{\diff \omega' \diff \Omega'} = 
  \int \frac{ \diff^3 p}{(2 \pi)^3 (2 \varepsilon)}\, \lvert \rho(\bm{p}) 
  \rvert^2 \, \frac{e^2 \omega'^2}{2(4\pi)^2 p^- q^-} \, \sum_{\sigma, \,\sigma', \, l} \lvert\overline{u}_{q,\sigma'} M_{fi}u_{p,\sigma}\rvert^2,
  \label{eq:incohaverage}
\end{equation}
where $q^{\mu}$ is a four vector such that $q^-=p^--k^{\prime,-}$,
$q_{x,y}=p_{x,y}-k'_{x,y}$ and
$q^+=(q^0+q^3)/2=(m^2+q_x^2+q_y^2)/2q^-$ ($q^2 = m^2$).  Equation
(\ref{eq:incohaverage}) can be easily identified as the incoherent
average over the modulus squared of $\rho(\bm{p})$ of the well-known
expression of the differential angular energy emission rate for a
nonlinear single Compton scattering event of an electron with definite
initial four-momentum $p^\mu$ and final four-momentum $q^\mu$
\cite{boca_nonlinear_2009, mackenroth2014quantum, felixnscs}
\begin{equation}
  \frac{\diff E_p}{\diff \omega' \, \diff\Omega'} =
  \frac{e^2 \omega'^2}{2(4\pi)^2\, p^-\, {q}^-} \sum_{\sigma,\, \sigma', l} \,\left\vert \overline{u}_{q,\sigma'}\,M_{fi}
    \, u_{p,\sigma} \right\vert^2.
  \label{eq:sing}
\end{equation}
Thus, there are no quantum interference effects between initial states
of the electron having different values of the momentum.  The physical
reason behind the absence of interference is that, in principle, by
measuring the final state of the electron and of the emitted photon
one can retrieve the initial momentum of the electron, and so the
initial state of the electron amongst the ones contained in the
initial superposition.

The results we presented so far allow us to state that, as far as one
is interested in nonlinear single Compton scattering rates, the state
of the initial electron can be described equivalently either with a
superposition of states like the one in \cref{eq:super} or as a
statistical mixture
\begin{equation}
  \hat{\rho}_\sigma = \int \frac{\diff^3 p}{ (2 \pi)^3 (2\varepsilon) } \, \lvert \rho(\bm{p}) \rvert^2 \, \ket{\Psi_{p, \sigma}}\bra{\Psi_{p, \sigma}}
\end{equation}
where the weighting function $\rho(\bm{p})$ is the same of
\cref{eq:super} and $\Psi_{p, \sigma}(x)=\langle x|\Psi_{p,
  \sigma}\rangle$.

\section{Electron wave-packets with normally distributed longitudinal
  momentum}\label{sec:diffpz}

After describing the theory for arbitrary superpositions of Volkov
states (for a given spin quantum number), in this section and in the
next one we will make an explicit choice of $\rho(\bm{p})$. Let the
initial state of the electron be a superposition of states with
momenta always directed almost in the opposite direction of the laser
wave-vector $\bm{k}$ (for the choice of the frame of reference we
adopted in \cref{sec:theo}, i.e., the momenta $\bm{p}$ are all
directed almost along the negative $z$ direction).  In particular, we
assume that the distribution of the momenta is a triple Gaussian
distribution, with average momentum
$\overline{\bm{p}}=(0,0,\overline{p}_z)$, with $\overline{p}_z<0$, and with
variance $\sigma_{p_T}^2$ along the $x$ and $y$ direction and $\sigma_{p_z}^2$
along the $z$-direction; thus the initial wave-packet is given by
\begin{equation}
  \Phi_\sigma(x) = \int \frac{\diff^3 p}{(2\pi)^3\, (2\varepsilon)} \frac{1}{\sigma_{p_T} \sqrt[4]{\sigma_{p_z}^2(2\pi)^3}}\,\, e^{-\frac{\left(p_z - \overline{p}_z\right)^2}{4 \sigma_{p_z}^2}} \,   e^{-\frac{p_x^2+p_y^2}{4 \sigma_{p_T}^2}}\, \Psi_{p,\sigma}(x).
  \label{Phi_parallel}
\end{equation}
In the present section the transverse variance $\sigma_{p_T}^2$ is assumed to be
sufficiently small, so that all transverse momenta $(p_x,p_y)$ in
Eq. (\ref{eq:incohaverage}) can be set equal to zero
(except than in the exponential in Eq. (22)). Thus, the
electron effectively collides head-on with the laser beam.

\begin{figure}
  \centering
  \includegraphics{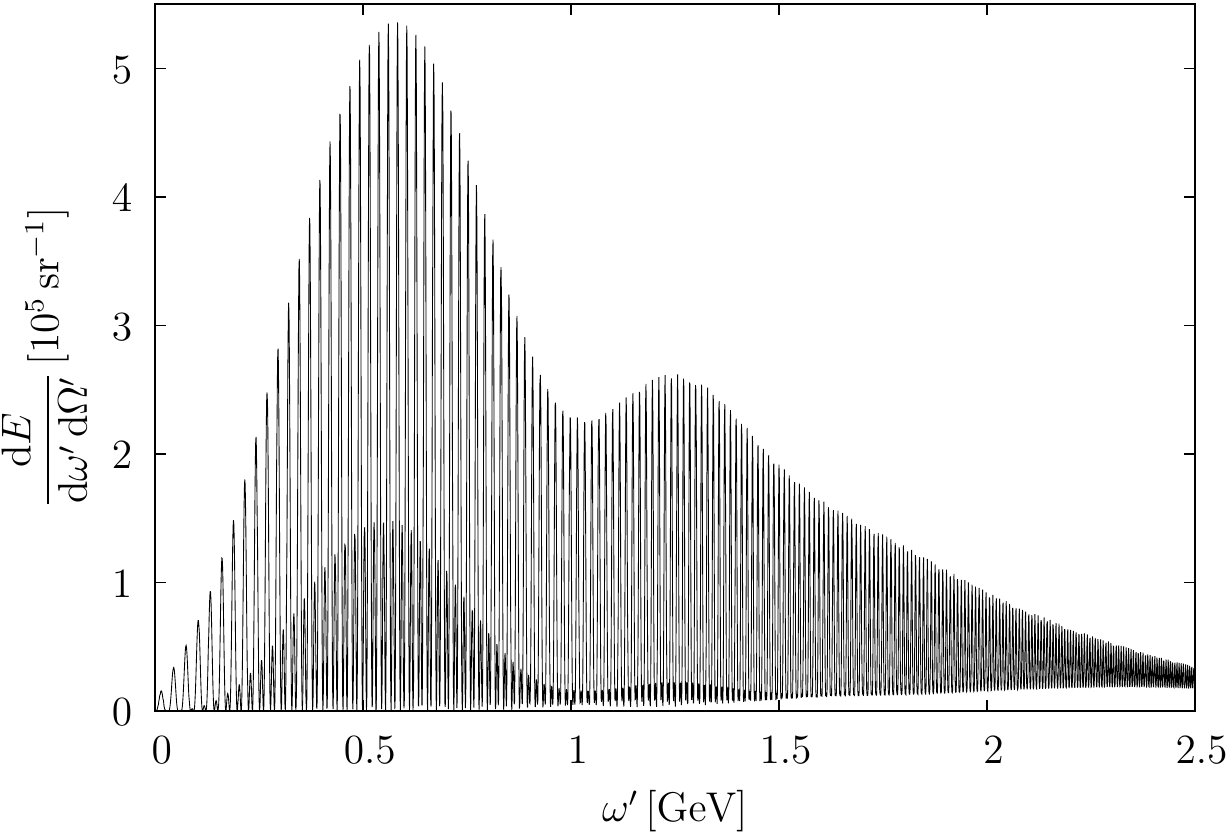}
  \caption{Energy emission spectrum along the negative $z$-direction
    for an incoming electron with definite initial momentum 
    $\bm{p}= (0,0,-4.2 \; \GeV)$ interacting with a laser
    of intensity $I \approx 4.3 \times 10^{20}\;\text{W}/\text{cm}^2$.}
  \label{fig:monochp3}
\end{figure}

\begin{figure}
  \includegraphics{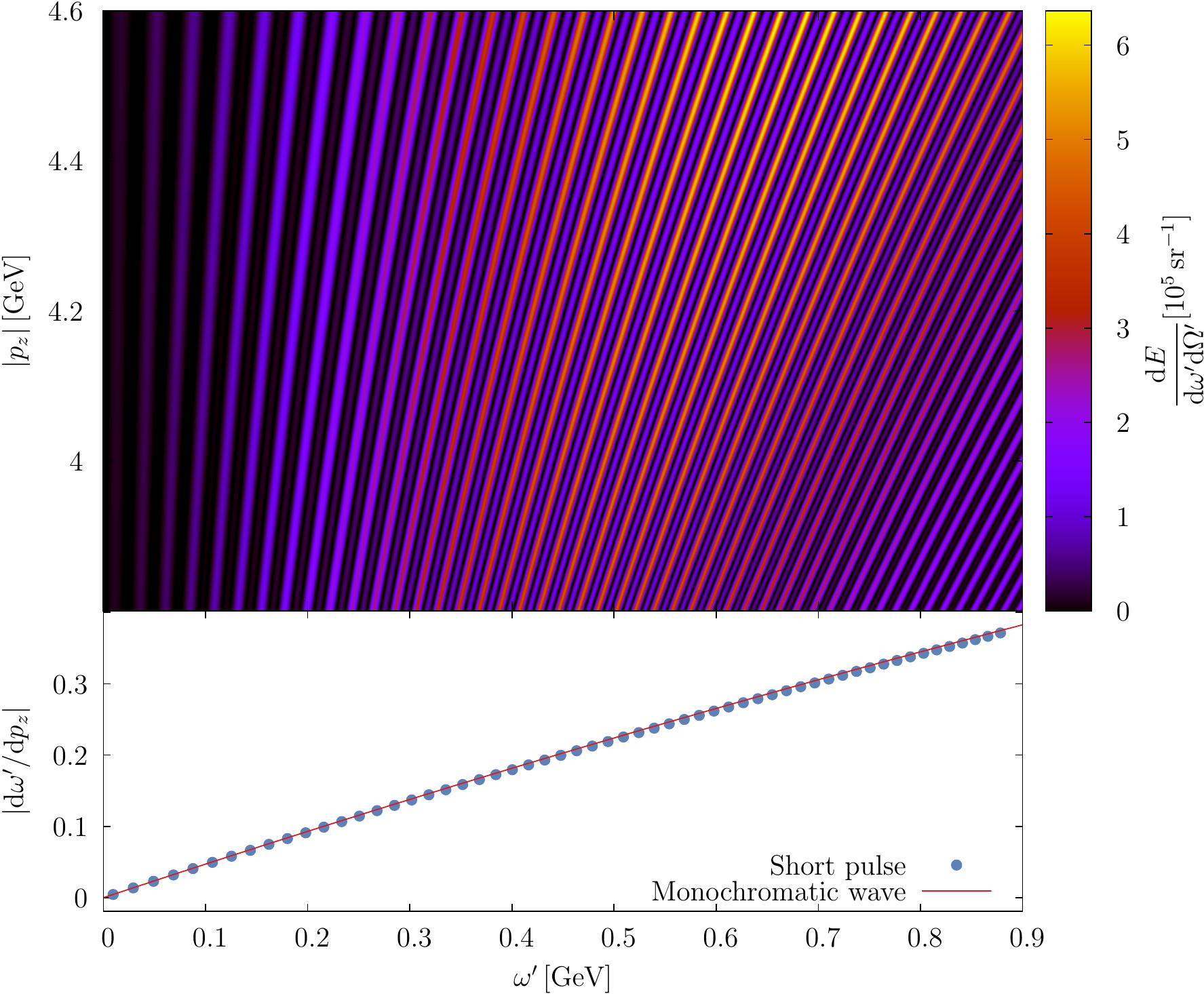}
  \caption{(color online) Change of the emission
    spectrum for an electron with definite initial momentum $(0,0,p_z)$
    as a function of $|p_z|$ (Fig. \ref{fig:monochp3} corresponds to a section of the upper part of
    this figure for $p_z = -4.2\;\GeV$). In the range considered, the
    position of the peaks increases linearly with $p_z$, albeit with
    different slopes depending on the position of the peak. 
    Some of these slopes were computed numerically
    and are shown in the bottom part of the plot (blue dots), together
    with the same quantity computed analytically for a monochromatic
    pulse (red continuous line).}
  \label{fig:long}
\end{figure}

In order to understand the modifications brought about by the electron
being described by the wave-packet in Eq. (\ref{Phi_parallel}), we
plot in \cref{fig:monochp3} the emission spectrum in the forward
(negative $z$) direction for an incoming electron with definite
momentum with components $p_x = p_y = 0$, and $p_z = -4.2\; \GeV$
($\varepsilon \approx 4.2\; \GeV$)~\cite{leemansbeam} interacting with a laser of
intensity $I \approx 4.3 \times 10^{20}\;
\text{W}/\text{cm}^2$. The above parameters
correspond to $\xi = 10$ and $\chi \approx 0.50$.
The spectra in the regime of $\lvert p_z \rvert \gg m$ and $\xi \gg
1$ exhibit a large number of narrow peaks. The position of the peaks
depends on the momentum of the electron; in particular, from
\cref{fig:long} one can deduce that, as the electron's initial
momentum increases in modulus, these peaks will be shifted towards
higher frequencies.
These shifts depend on the position of the peaks itself, i.e.,
different peaks are shifted by a different amount, when changing
$p_z$. More specifically, by changing $p_z$ of the same amount, the
higher peak frequencies will be shifted more than the lower ones. The
above results can be easily explained as a result of the Doppler
effect. For the sake of simplicity we consider here the idealized case
of a monochromatic laser field (with laser photon energy $\omega$). 
In this case, in fact, the frequency of the $n^\text{th}$
harmonic emission along the negative $z$ direction is given by
\citep{ritus}
\begin{equation}
  \omega_n' = \frac{n(p k)}{(p n') + \left(n + \frac{m^2 \xi^2}{4 (p k)} \right)(k n')} = \frac{n \omega \left(\varepsilon -
      p_z\right)^2}{m^2 \left( 1 + \frac{\xi^2}{2}\right) + 2 n \omega (\varepsilon - p_z)}=\frac{\zeta_n}{1+2\zeta_n}(\varepsilon-p_z),
  \label{eq:armoniche}
\end{equation}
where $n'^\mu = (1, \bm{k}'/\omega')=(1,0,0,-1)$ and where we have
introduced the dimensionless parameter
\begin{equation}
  \zeta_n = \frac{n \omega (\varepsilon - p_z)}{m^2 \left( 1 + \frac{\xi^2}{2}\right)}.
  \label{eq:zetadef}
\end{equation}
By means of a first-order expansion with respect to the shift $\Delta
p_z$, we can estimate the relative shift of these frequencies when
slightly changing the value of $p_z$:
\begin{equation}
  \frac{\Delta\omega_n'}{\omega_n'} = \frac{1}{\omega_n'} \, \frac{\partial
    \omega_n'} {\partial p_z} \, \Delta p_z = -2\, \frac{1+\zeta_n}{1+2\zeta_n}
  \, \frac{\Delta p_z}{\varepsilon}.
  \label{eq:varipz}
\end{equation}
In the case of an ultrarelativistic electron and in the relevant
regime $\xi\gg 1$, it is $\varepsilon \approx \left| p_z\right|$ and
$\zeta_n\approx 2n\chi/\xi^3$, such that we obtain
\begin{equation}
  \frac{\Delta\omega_n'}{\Delta |p_z|} \approx 4\, \frac{\zeta_n(1+\zeta_n)}{(1+2\zeta_n)^2}.
  \label{eq:shiftfreq}
\end{equation}
As it can be easily shown, the quantity $\Delta\omega_n'/\Delta |p_z|$
increases monotonically with the harmonic number, in agreement with
the findings in Fig. \ref{fig:long}.

Notice that \cref{eq:shiftfreq} is valid only for a monochromatic
laser field, whereas we are interested here in the case of short
pulses, i.e., pulses also characterized by a certain spread
$\Delta\omega$ around a central angular frequency $\omega$. It is thus
interesting to compare the relative shift due to an uncertainty of
$p_z$ to the one due to an indeterminacy on the value of $\omega$.  In
analogy to what we have discussed for~\cref{eq:varipz}, one can derive
a similar relation, for a variation $\Delta\omega$ of the laser
angular frequency. By adding the resulting expression to
\cref{eq:varipz} and by assuming again that $|\bar{p}_z| \gg m$ and $\xi\gg
1$, it is possible to obtain the first-order relative variation of
$\omega_n'$ with respect to the relative variations of $\omega$ and
$p_z$ as:
\begin{equation}
  \frac{\Delta\omega_n'}{\omega_n'} \approx \frac{1}{1+2\zeta_n}\, \frac{\Delta \omega}
  {\omega} + 2\, \frac{1+\zeta_n}{1+2\zeta_n}
  \, \frac{\Delta |p_z|}{\left| p_z\right|}.
  \label{eq:shiftmono1}
\end{equation}
Since $\zeta_n>0$ it is clear that for comparable relative variations in
$\omega$ and $p_z$, the induced shift due to the spread in the
incoming electron momentum is larger. From the aforementioned properties of the emitted photon's spectrum of
a monochromatic initial electron we can infer the final spectrum when
the state $\Phi_\sigma(x)$ of \cref{eq:super} is considered, since the
emission spectrum resulting from that state, as it was shown above, is
a weighted average of monochromatic emission spectra with different
$p_z$. The sharp peaks present in the spectrum for a fixed value of
$p_z$ will be differently shifted and will tend to fill the valleys
present in the spectrum relative to another value of $p_z$; when
averaging many of these spectra, the net effect is a smoothing of the
final spectra and a decrease of the yield as compared to the latter
obtained at the peaks in the monochromatic case.

\begin{figure}
  \includegraphics{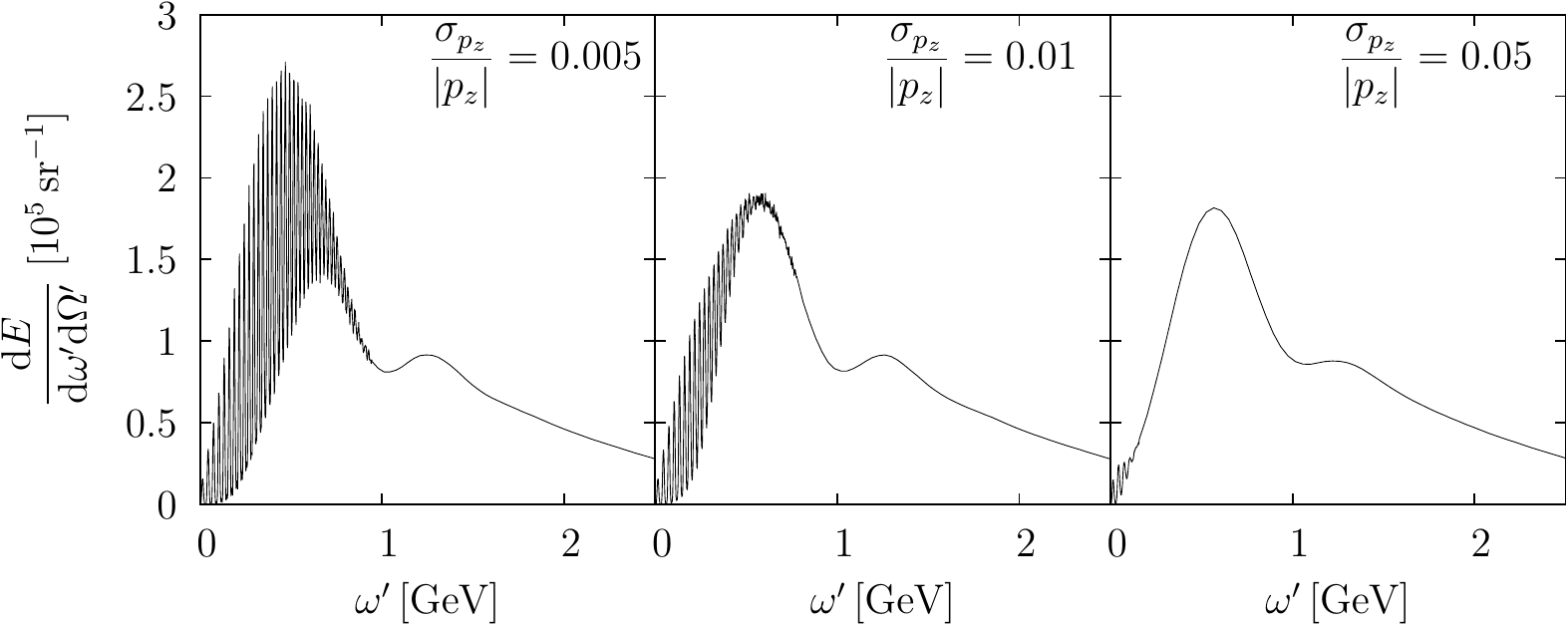}
  \caption{Emission spectra along the negative $z$-direction for an
    electron wave-packet with $\bar{\bm{p}}=(0,0,-4.2\;\text{GeV})$ 
    interacting with a laser pulse of peak intensity $I \approx 4.3 \times 10^{20}\;
    \text{W}/\text{cm}^2$ for different
    values of the spread of the longitudinal momentum.}
  \label{fig:dist}
\end{figure}

Moreover, we have already mentioned the fact that the shift induced by
the spread in the electron momentum is larger for higher emission
frequencies. Thus, the portion of the spectrum that will be smoothed
earlier, i.e., even for relatively small values of $\sigma_{p_z}$, is that
at high frequencies of the emitted photon.  Indeed, this is the result
we obtain in \cref{fig:dist}, where the final photon energy spectrum
for different values of $\sigma_{p_z}$ is plotted (the numerical parameters are the
same as in \cref{fig:monochp3} and the average value of the initial momentum
of the electron is $\bar{\bm{p}}=(0,0,-4.2\;\text{GeV})$).
We have chosen values of the standard deviation $\sigma_{p_z}$ equal to $0.5\%$,
$1\%$ or $5\%$ of the incoming momentum, corresponding to $21\; \MeV$,
$42\; \MeV$ or $210\; \MeV$, respectively. Even when the relative
indeterminacy on the momentum is only $0.5\%$, we can see that the
height of the highest peaks is reduced by a factor of about two, and
all the oscillatory features at $\omega' \gtrsim 1\; \GeV$ are
completely washed out (see Fig. \ref{fig:monochp3} and Fig. \ref{fig:dist}). 
For larger values of $\sigma_{p_z}$, these effects
are even more evident also for the lowest part of the spectrum. Concerning
the choice of $\sigma_{p_z}$ and in general of the properties
of the wave-packet $\Phi_\sigma(x)$ a comment is in order. In fact, in general,
the state $\Phi_\sigma(x)$ describes a single electron. The
properties of the corresponding wave-packet depend on how
the electron is produced and accelerated~\cite{Baum_2013},
and are in principle different from, for example, the
corresponding properties of an electron bunch. However, in our
case, as we have seen, the spectra for
the state $\Phi_\sigma(x)$ coincide with those
obtained by considering a corresponding electron bunch with an
average electron number equal to unity. In this
respect, the values of the momenta spreads are chosen
according to the features of electron beams, which
can be obtained presently experimentally~\cite{leemansbeam}.

\section{Multivariate Gaussian wave-packets}\label{sec:multi}

We now turn our attention to the experimentally more realistic
situation of an electron wave-packet that can have also non-zero
components of the transverse momentum. Our choice for the initial
state is as in Eq. (\ref{Phi_parallel}) but this time the variance
$\sigma_{p_T}^2$ is assumed not to be small. Also in this case, as we did in the previous section, we will first
consider how the spectrum of electrons initially in a Volkov state in
a monochromatic field is modified as a function of the components of
the initial momentum. Then, starting from those considerations, we
will focus onto the case of an electron wave-packet in a short laser
pulse.

\begin{figure}
  \includegraphics{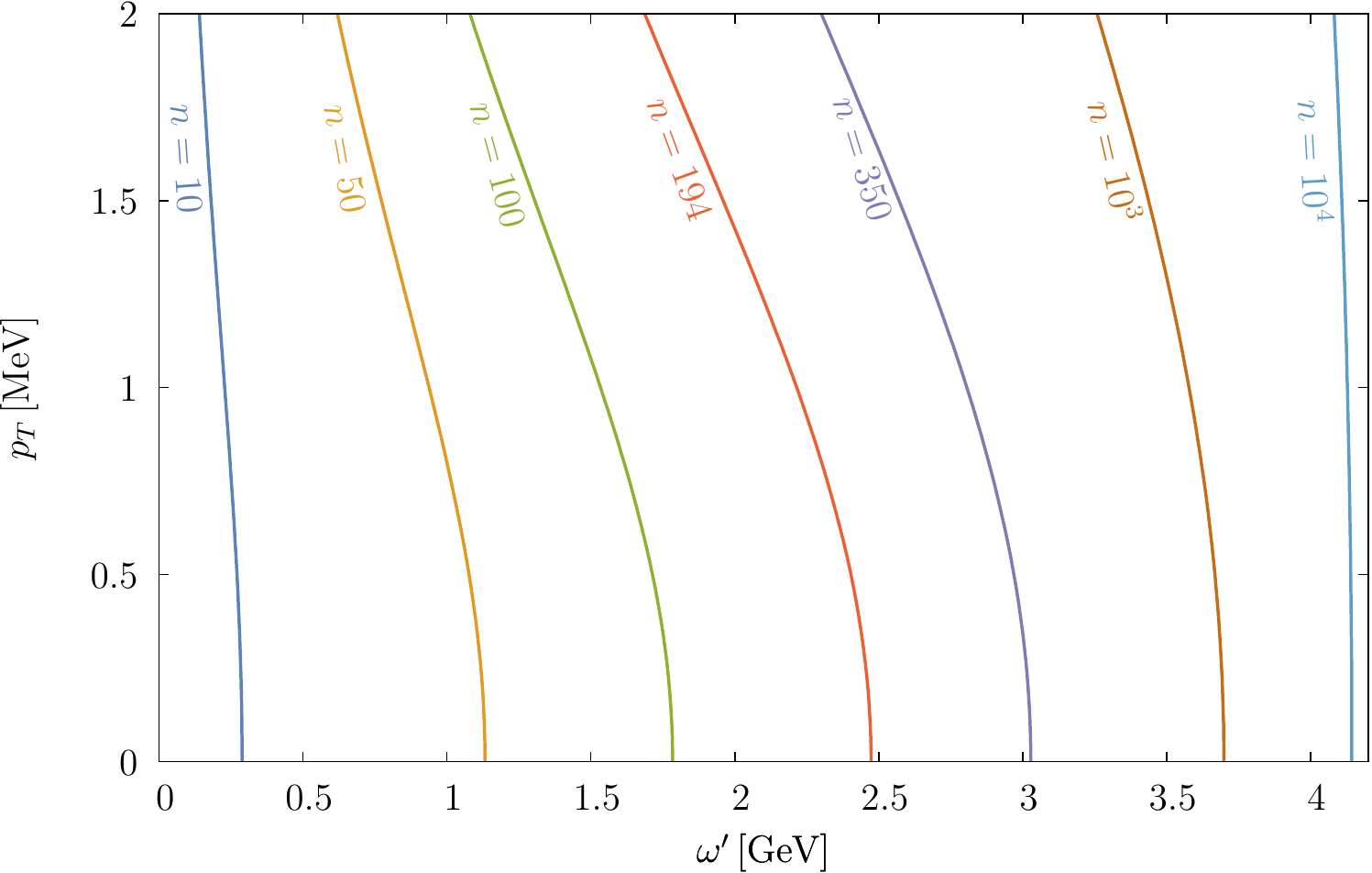}
  \caption{(color online) Shift of the emission frequencies
    $\omega_n'$ along the negative $z$-direction for different values
    of $n$ as a function of $p_T$ (vertical axis). The numerical
    parameters are $p_z = - 4.2 \; \GeV$ and $I \approx 1.1 \times 10^{20} \;
    \text{W}/\text{cm}^2$.}
  \label{fig:ptshift}
\end{figure}

In order to understand how the emission spectrum is altered by the
possibly non-zero value of the transverse components of the initial
momentum, we show how the harmonic frequencies along the negative
$z$-direction are shifted as the transverse momentum $p_T =
\sqrt{p_x^2 +p_y^2}$ varies.  We can thus proceed in analogy to the
derivation of~\cref{eq:shiftfreq}. The starting point is the initial
form of $\omega_n'$ in \cref{eq:armoniche}, which can be rewritten in
the more convenient form
\begin{equation}
  \omega_n' = \frac{n \omega \left(\varepsilon -
      p_z \right)^2}{m^2 \left( 1 + \frac{\xi^2}{2}\right)+ p_T^2  + 2 n \omega (\varepsilon - p_z)},
\end{equation}
showing the explicit dependence also on $p_T^2$ (remember that now
also the energy $\varepsilon$ depends on $p_T^2$). By expanding
$\omega_n'$ around $p_T = 0$ we obtain
\begin{equation}
  \frac{\Delta \omega_n'}{\omega_n'} = \frac{1-(\varepsilon/n\omega-1)\zeta_n}{1+2\zeta_n}\frac{\Delta p_T^2}{\varepsilon(\varepsilon-p_z)},
\end{equation}
where all the energies are calculated at $p_T=0$. This equation shows
that again the relative shift depends on the harmonic number $n$. In a
typical scenario where $\varepsilon\approx |p_z|$ and $\xi\gg 1$, the
same approximations as in the previous section can be applied. The
result for $\Delta\omega_n'$ reads
\begin{equation}
  \Delta \omega_n' = \zeta_n\frac{1+\zeta_n-\varepsilon\chi/\xi^3\omega}{(1+2\zeta_n)^2}\frac{\Delta p_T^2}{\varepsilon},
\label{eq:Deltaomeganprime}
\end{equation}
with $\zeta_n$ given in~\cref{eq:zetadef}, which in the current 
approximations ($\varepsilon \approx |p_z|$, $\xi \gg 1$) is approximately equal
to $n\chi/\xi^3$. \Cref{eq:Deltaomeganprime} shows that an important role is played
by the parameter $\mu=\varepsilon\chi/\xi^3\omega$.  If we work in
the quantum regime where $\chi\sim 1$, since at $\xi\sim 10^2$
electron energies in the GeV-range are required, we can safely assume
that $\mu\gg 1$. Moreover, at $\zeta_n\gg 1$
the emission spectrum is suppressed~\cite{ritus} such that we can conveniently
further approximate the expression for $\Delta \omega_n'$ as
\begin{equation}
  \Delta \omega_n' = -\frac{\varepsilon\chi}{\xi^3\omega}\frac{\zeta_n}{(1+2\zeta_n)^2}\frac{\Delta p_T^2}{\varepsilon}.
  \label{delta_omegap}
\end{equation}
This expression indicates that we would expect a negative shift of the
harmonics, which becomes less pronounced at $\zeta_n\ll 1$ (low
harmonics) and at $\zeta_n\gg 1$ (high harmonics). This is exactly
what we observe in \cref{fig:ptshift}, where different curves
$\omega_n'=\omega_n'(p_T)$ for different values of $n$ are plotted, with the
numerical parameters: $p_z = - 4.2 \; \GeV$ and $I \approx 1.1 \times 10^{20} \;
\text{W}/\text{cm}^2$ ($\xi = 5$, $\chi \approx 0.25$, and
$\mu\approx 5.4\times 10^6$).

\begin{figure}
  \includegraphics{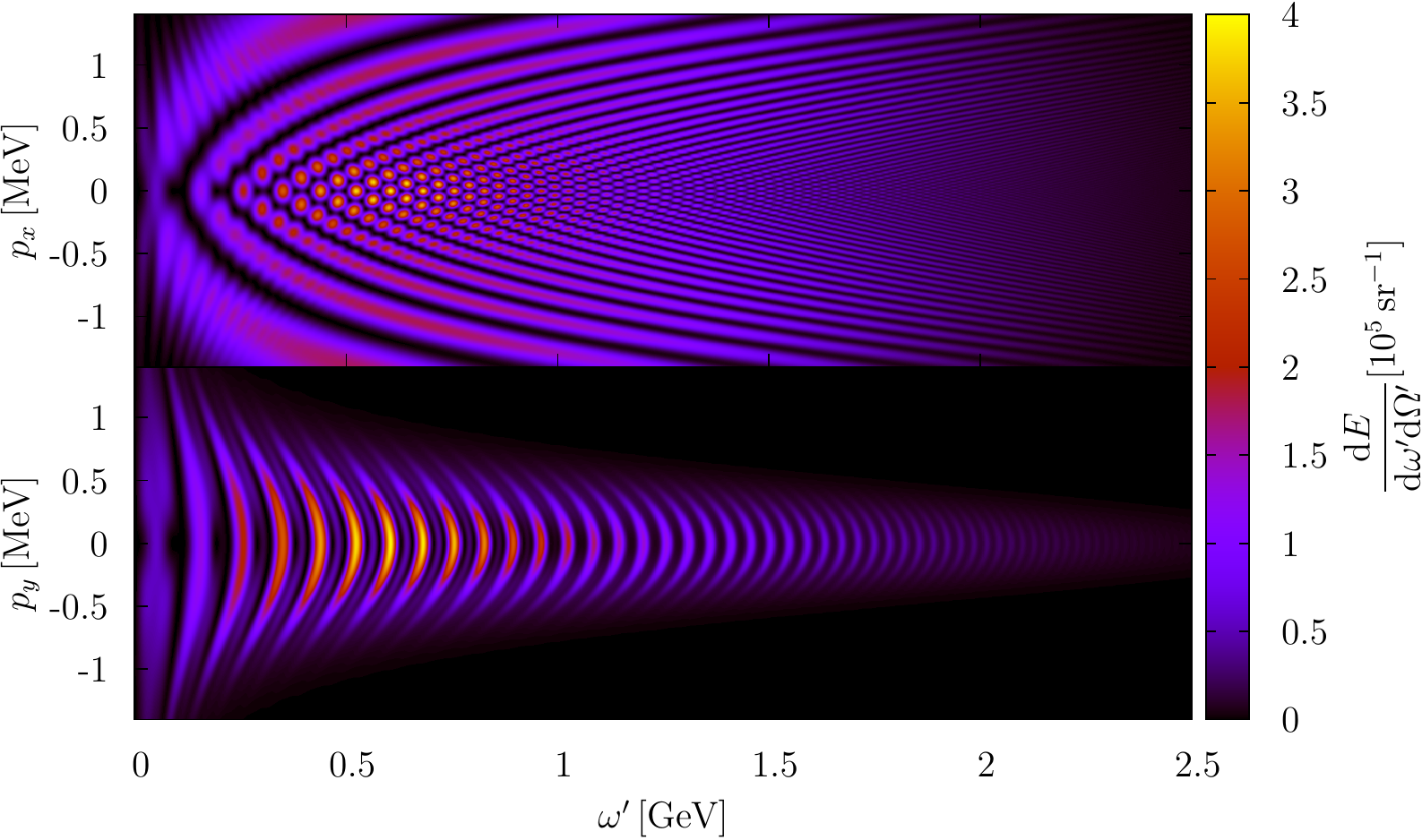}
  \caption{(color online) Emission spectra in the negative $z$-direction
    for electrons having initially $p_z = -4.2\; \GeV$ and either $p_y = 0$ or $p_x = 0$,
    after the interaction with a short laser pulse with $I \approx  1.1 \times 10^{20} \;
    \text{W}/\text{cm}^2$.
  }
  \label{fig:sppx}
\end{figure}

A typical collection of monochromatic electron spectra along the
forward direction is shown in \cref{fig:sppx} by electrons having
initially $p_z = -4.2\; \GeV$ and either $p_y = 0$ or $p_x = 0$
(we recall that $p_x$ ($p_y$) is the component of the
momentum along the direction of the electric (magnetic)
field of the laser), interacting with a short laser pulse with 
$I \approx  1.1 \times 10^{20} \;
\text{W}/\text{cm}^2$ ($\xi = 5$, $\chi \approx 0.25$).
Apart from exhibiting the already mentioned shift of the peak
frequencies as one of the transverse components varies, we also observe that by
varying $p_y$ by about 1-2 electron masses the spectrum is
significantly suppressed. The reason is that the observation
direction is the forward direction and that the angular emission
range of the electron along the magnetic field of the laser is of
about $m/\varepsilon$, whereas along the electric field of the laser,
the electron emits up to angles of the order of $m\xi/\varepsilon$~\cite{mackenroth2014quantum}.
It is also worth observing the large oscillations in the emitted
intensity between successive peaks when varying $p_x$ (top part of
\cref{fig:sppx}). These oscillations are expected to have an important
effect, when averaging many spectra, even for $|p_x| \ll m\xi$.

\begin{figure}
  \includegraphics{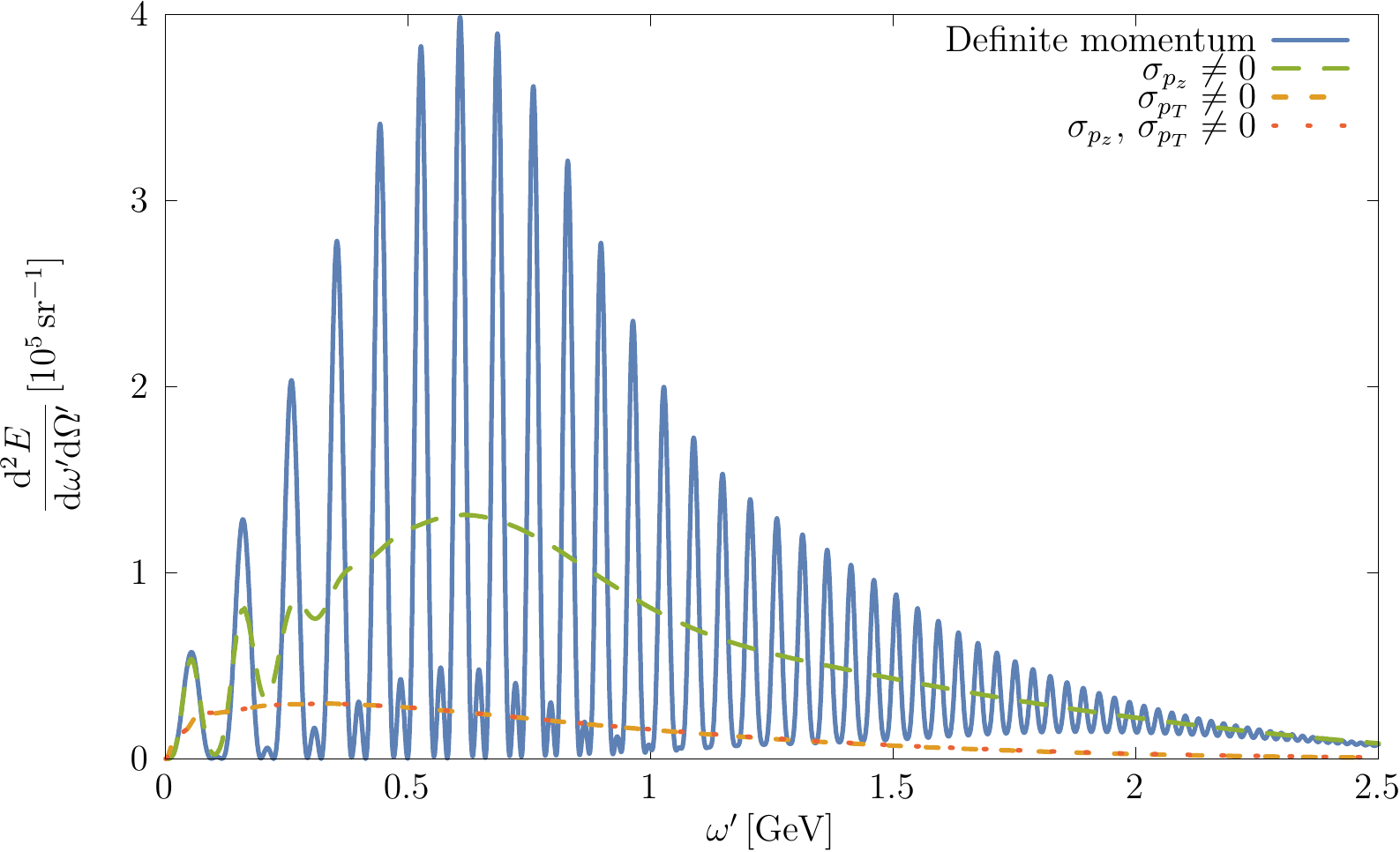}
  \caption{(color online) Energy emission spectrum in the negative $z$-direction, 
    for some different initial electron states. 
    Here, $\bar{\bm{p}} = (0,0,-4.2\; \GeV)$, $\sigma_{p_T} = 3 \times 10^{-4} 
    \, \lvert \bar{\bm{p}} \rvert$, and $\sigma_{p_z}= 6 \times 10^{-2} \, 
    \lvert \bar{\bm{p}} \rvert$. The intensity of the laser field is $I \approx 1.1 \times
    10^{20} \; \text{W}/\text{cm}^2$.
  }
  \label{fig:distzpos}
\end{figure}

\begin{figure}
  \includegraphics{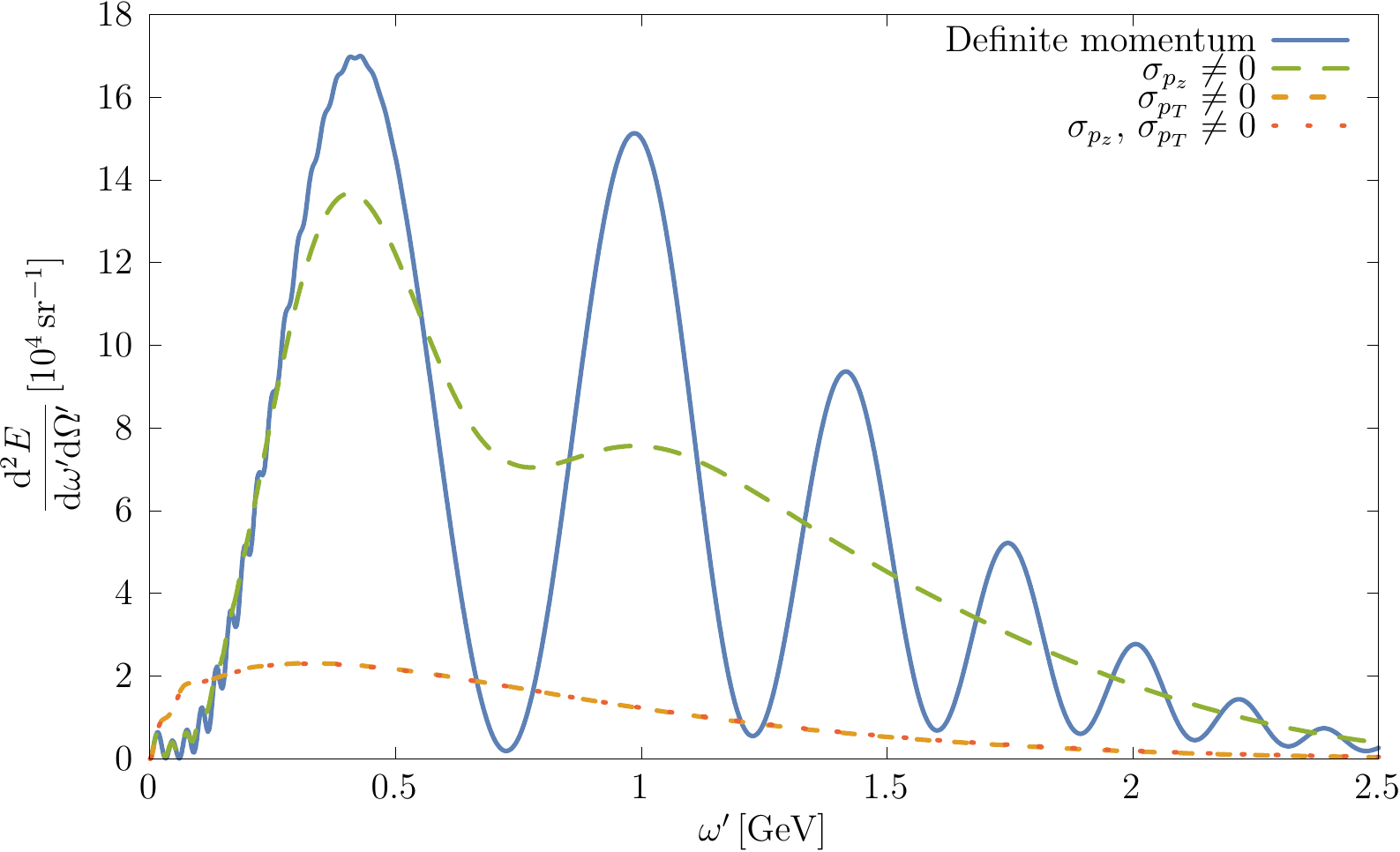}
  \caption{(color online) Energy emission spectrum on a direction in
    the $xz$-plane forming an angle $\theta= m\xi/2\bar{\varepsilon}$ with
    the negative $z$-axis, for some different initial electron
    states. The numerical parameters are the same as in \cref{fig:distzpos}.}
  \label{fig:distang}
\end{figure}

The above observations are confirmed by numerical calculations. In \cref{fig:distzpos} 
(\cref{fig:distang}), we show the effects on the spectrum of the emitted photon 
along the negative $z$ direction (in a direction that lies on the $xz$-plane, the laser 
polarization-propagation plane, and forms an angle $\theta
= m\xi/2\bar{\varepsilon}$ with the negative $z$-axis, where 
$\bar{\varepsilon}$ is the average initial electron energy) of having 
either $\sigma_{p_T}\neq 0$ or $\sigma_{p_z}\neq 0$, or 
$\sigma_{p_T},\sigma_{p_z}\neq 0$ (in the first two cases $\sigma_{p_z}$ 
and $\sigma_{p_T}$, respectively, are considered to be sufficiently small 
that their effects can be neglected as explained below Eq. (\ref{Phi_parallel})).

In the numerical spectra in \cref{fig:distzpos} and \cref{fig:distang}, the average 
initial momentum of the electron
is $\bar{\bm{p}} = (0,0,-4.2\; \GeV)$, and the indeterminacy on the
transverse components is $\sigma_{p_T} = 3 \times 10^{-4} \, \lvert
\bar{\bm{p}} \rvert$, while the one on the $z$-component is $\sigma_{p_z}
= 6 \times 10^{-2} \, \lvert \bar{\bm{p}} \rvert$ (these parameters
for the electron beam are compatible with those in
\cite{leemansbeam}). The intensity of the laser field is $I \approx 1.1 \times
10^{20} \; \text{W}/\text{cm}^2$ ($\xi = 5$, $\chi=\bar{\chi} \approx
0.25$ as calculated from the average electron momentum).

In \cref{fig:distzpos} and \cref{fig:distang} one can see that for the
chosen values of the parameters $\sigma_{p_T}$ and $\sigma_{p_z}$, the most dramatic
alteration of the spectrum is due to the transverse momentum spread of
the electron beam, even though its value is orders of magnitude
smaller than the spread on $p_z$. In fact, the effect due to $\sigma_{p_T}
\neq 0$ is so dominant that switching on also the longitudinal spread
$\sigma_{p_z}$ has no observable effect on the emitted spectrum (the
dotted red curve is on top of the short-dashed orange one in both \cref{fig:distzpos}
and \cref{fig:distang}).  As a result, the finer structures in the
spectra are washed out and, in this respect, in order to at least
partially observe them one should experimentally render the incoming
electron beam as collimated as possible.
\begin{figure}
  \includegraphics{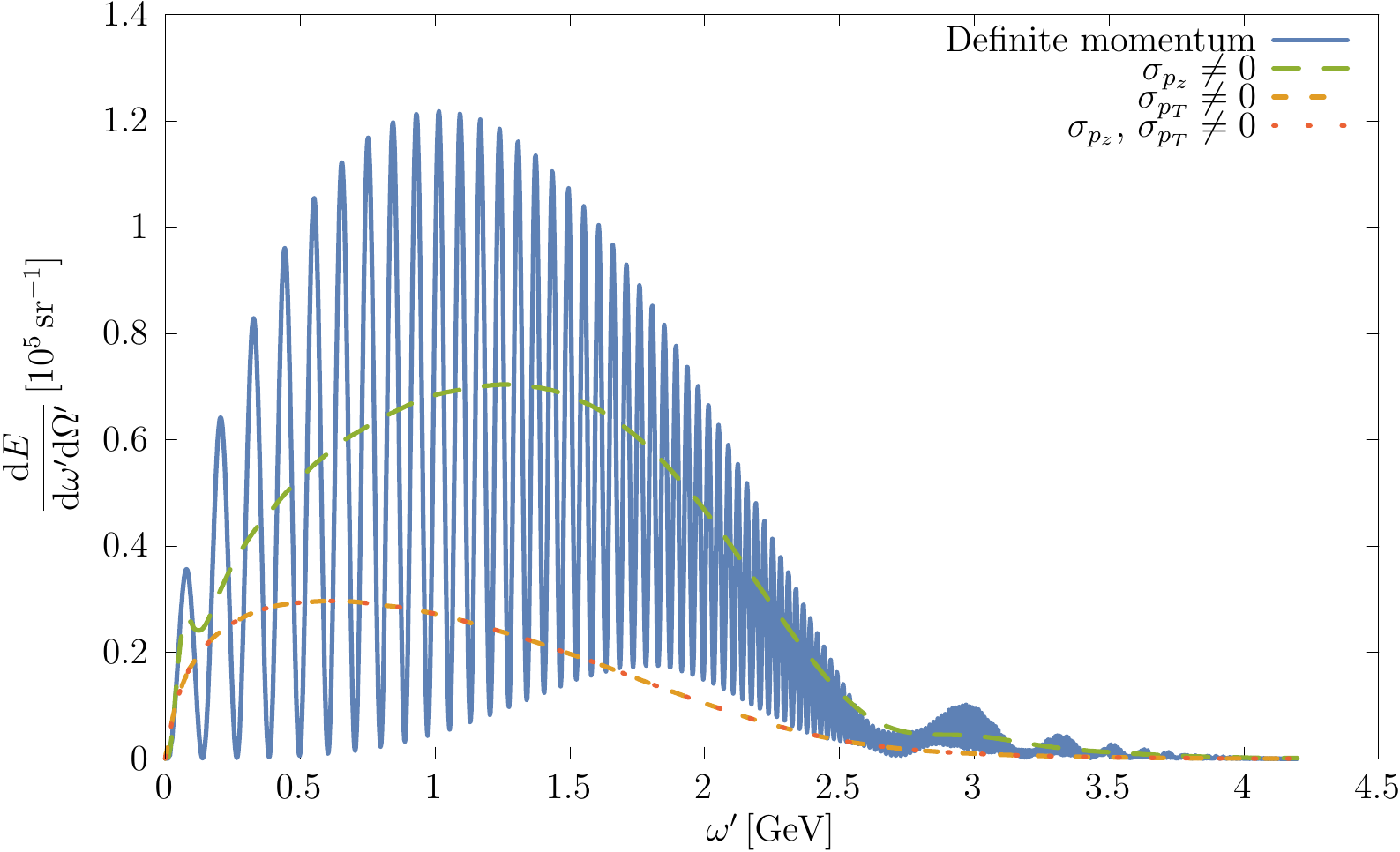}
  \caption{(color online) Energy emission spectrum for an electron
    wave-packet in the quantum regime ($\chi \approx 0.85$), in the
    direction that lies on the laser polarization plane and forms an
    angle $m\xi/2\bar{\varepsilon}$ with the negative z-axis. 
    The numerical parameters are the same as in \cref{fig:distzpos}, 
    except that $I \approx 1.2 \times 10^{21} \; \text{W}/\text{cm}^2$.}
  \label{fig:hixi}
\end{figure}

\begin{figure}
  \centering
  \includegraphics{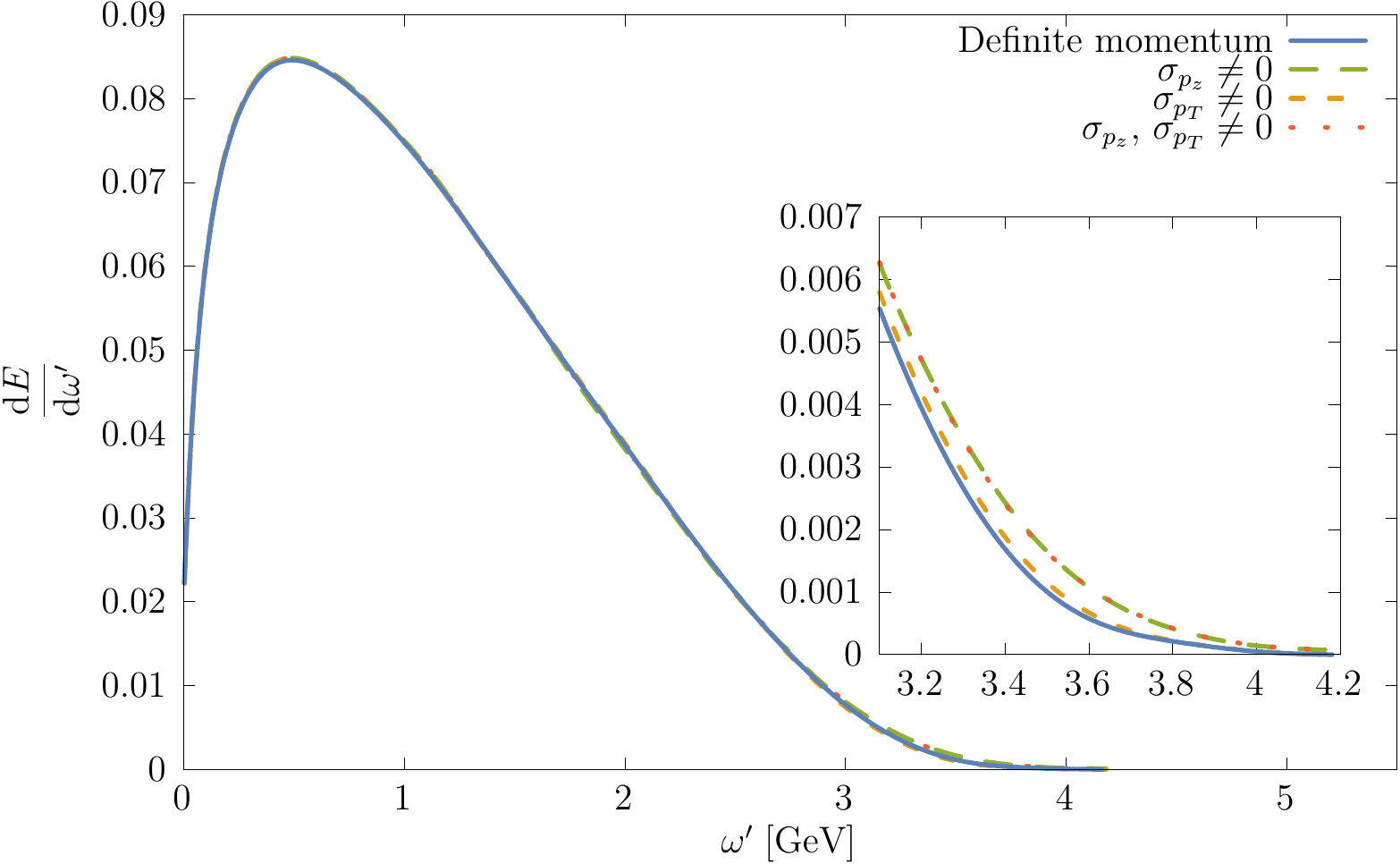}
  \caption{(color online) Distribution of the total emitted energy by 
    an electron in a Volkov state or in a Gaussian wave-packet as a 
    function of the frequency of the emitted photon. All the numerical 
    parameters for this figure are the same as in \cref{fig:hixi}.}
  \label{fig:angintegrated}
\end{figure}

We also show in \cref{fig:hixi} the energy emission along a direction
that lies on the laser polarization plane and forms an angle
$m\xi/2\bar{\varepsilon}$ with the negative z-axis for 
$\chi=\bar{\chi}\approx 0.85$ (the parameters used for \cref{fig:hixi} are the same
of \cref{fig:distang}, except that $I \approx 1.2 \times 10^{21} \;
\text{W}/\text{cm}^2$ corresponding to $\xi = 17$); the qualitative
behavior for nonzero values of $\sigma_{p_z}$ and $\sigma_{p_T}$ is the
the same as the one previously discussed. We should emphasize that,
as we have already mentioned in the discussion below Eq. (\ref{delta_omegap}),
the larger effect due to the transverse momentum uncertainty is also
related to the fact that the considered spectra refer to some
specific observation directions. In fact, 
if we integrate with respect to the emission angles the spectrum 
corresponding to the numerical parameters in \cref{fig:hixi}, we obtain
the results in \cref{fig:angintegrated}; they show that the total
emitted energy as a function of $\omega'$ changes only at frequencies
$\omega' \approx \bar{\varepsilon}=4.2\;\text{GeV}$ and that it is
almost not affected by the momentum spreading of the incoming wave-packet. 
The higher rates observed at these frequencies in the case of a
wave-packet with $\sigma_{p_z}\neq 0$ (see inset of \cref{fig:angintegrated}) can be explained 
as some components of the wave-packet have energies larger than 
$\bar{\varepsilon}$.

\begin{figure}
  \centering
  \includegraphics[width=.94\textwidth]{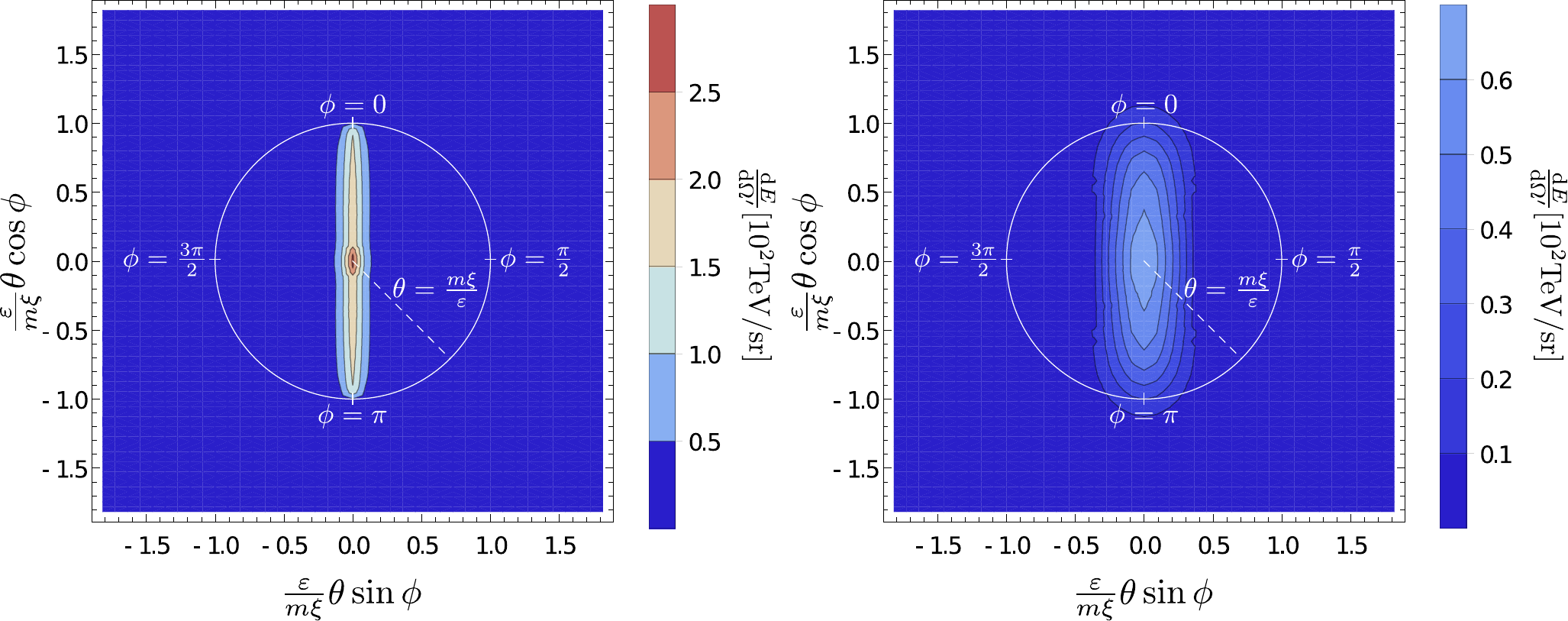}
  \caption{(color online) Angular distribution of the total energy emitted 
    by an electron in a Volkov state (left) or in a Gaussian superposition 
    of them (right) after interacting with a strong laser field. The numerical 
    parameters used here are the same as in \cref{fig:hixi}.}
  \label{fig:total}
\end{figure}

In order to analyze the properties of the emitted radiation in the
spatial domain, one can integrate $ \diff E/\diff \omega' \,
\diff\Omega'$ with respect to $\omega'$ and obtain the total
energy emitted along each direction.  A typical result of this procedure
is shown in \cref{fig:total}. On the right panel the energy
emitted per steradian by an electron in a Gaussian wave-packet is plotted 
(the numerical parameters are the same as in \cref{fig:hixi}). The left panel 
shows the same quantity but emitted by an electron in a Volkov state 
with a definite momentum given by the $\bar{\bm{p}}$ of the mentioned Gaussian 
wave-packet. In \cref{fig:total} the polar angle $\theta$ and the azimuthal 
angle $\phi$ are indicated assuming the negative $z$-axis as polar axis.
As mentioned above, when the electron is initially in a pure Volkov state, and the laser is
linearly polarized, the angular aperture of the emitted radiation is $m\xi/\varepsilon$ ($m/\varepsilon$)
along the polarization (magnetic-field) direction, which is confirmed by the
the left panel in \cref{fig:total}. The emission in the case of a multivariate Gaussian
wave-packet, in the right panel of \cref{fig:total}, extends over a broader
region and is thus less intense, in the regime where $\sigma_{p_T}$ and
$\sigma_{p_z}$ are much smaller than $|\bar{p}_z| \gg m$. In fact, at 
$\xi \gg 1$, if $\sigma_{p_T} \ll |\bar{p}_z|$ and $\sigma_{p_z} \ll |\bar{p}_z|$, the total energy
emitted when the electron is either in a Volkov state or in a Gaussian
wave-packet is almost the same (see Fig. \ref{fig:angintegrated}). Then, as the region of emission becomes 
broader, the radiation intensity in the Gaussian wave-packet case
decreases. We briefly notice here that this effect might be also exploited
in principle as a diagnostic tool of the momentum spreading of the
electron beam, provided that the laser parameters like its intensity
are known with sufficiently high accuracy.

\section{Conclusions}

In the present article we have studied nonlinear single Compton
scattering by an incoming electron described by a wave-packet of
Volkov states. We have obtained that the conservation of energy and
momentum forbids interference effects among different momentum
components of the wave-packet, even if the electron is
originally in a superposition of Volkov states. This means that an incoming
electron wave-packet can be equivalently described in this respect as
a superposition of states or as a statistical mixture. The net effect
of having a wave-packet as initial electron state is a lowering and a
smoothing of the angular resolved emission spectrum for an electron in a state with
definite momentum; this effect tends to be more pronounced than the
non-monochromaticity of the laser pulse (at comparable relative
uncertainties in the electron and in the laser-photon
energy). Furthermore, for realistic values of the properties of the
electron wave-packet as compared with those available experimentally
for electron beams, the transverse momentum spread, even if orders of
magnitude smaller than the longitudinal one, dominates the alterations
on the structures and on the shape of the emission spectrum at a fixed
observation direction. We have observed that a broadening of the angular 
emission region takes also place in the case of an electron wave-packet
with respect to the case of a monoenergetic electron. However, by integrating
the spectra over the observation directions, their dependence on the
spreading of the initial wave packet is strongly suppressed.

\bibliographystyle{unsrt} 

\end{document}